\documentclass{jaa}

\usepackage{graphicx}
\usepackage{subcaption}
\usepackage{xcolor}

\begin{document}\sloppy

\title {Bursts of Gravitational Waves due to Crustquake from Pulsars}


\author{Biswanath Layek\textsuperscript{1,*} and Pradeepkumar Yadav\textsuperscript{1}}
\affilOne{\textsuperscript{1}Birla Institute of Technology and Science, Pilani Campus \\ 
Pilani, Jhunjhunu 333031, Rajasthan, India\\}


\twocolumn[{

\maketitle

\corres{layek@pilani.bits-pilani.ac.in}


\begin{abstract}
We revisit here a possibility of generation of gravitational wave (GW) bursts due to a very quick
change in the quadrupole moment (QM) of a deformed spheroidal pulsar as a result of crustquake.
Since it was originally proposed as a possible explanation for sudden spin-up (glitch) of pulsars, 
the occurrence of crustquake and it's various consequences have been studied and discussed quite 
often in the literature.  Encouraged by recent development in gravitational wave (GW) astronomy, we re-investigate 
the role of crustquake in the emission of GWs. Assuming exponential decay of excitation caused by crustquake, 
we have performed a Fourier analysis to estimate the GW strain 
amplitude $h(t)$, characteristic signal amplitude $h_c(f)$ and signal to noise ratio (SNR) of the burst for
the Crab pulsar. For exotic quark stars, multifold enhancement of these quantities are expected, which 
might make quark star a potential source of gravitational waves. The absence of such bursts may put several 
constraints on pulsars and such hypothetical stars. 
\end{abstract}

}]


\doinum{12.3456/s78910-011-012-3}
\artcitid{\#\#\#\#}
\volnum{000}
\year{0000}
\pgrange{1--}
\setcounter{page}{1}
\lp{1}

\section{Introduction}
The remarkable first-ever direct detection of GWs in 2015 by LIGO (Abbott \& {\em et al.} 2016)
opened a new era in gravitational astronomy. Observations supplemented with numerical simulations
identified a black hole merger as the source for \lq ripples' in the space-time. Since then there 
have been quite a few significant detections of GWs. The peak strain amplitude ($h$) 
for all these detections have been in the range $10^{-21}~-~10^{-22}$. In view of these detections, 
we should be hopeful that the gravitational waves produced by compact isolated astrophysical objects like 
isolated pulsars can be measurable in the near future by more sensitive upcoming ground-based advanced detectors, 
namely, aLIGO, VIRGO, the third generation Einstein Telescope (ET), etc. Prior to the GWs detections as 
mentioned above, there have been a few attempts for GWs searches from isolated pulsars as well. 
The attempt for GWs search associated with the timing glitch in the Vela pulsar 
in August 2006 (Abadie {\em et al.} 2011) was one among those which are worth mentioning. The motivation of the above 
searches were based on suggestions made by several authors (see the Ref. Abadie {\em et al.} (2011) and 
the list of references therein) through their works in this area. As per those suggestions, several sources 
namely the flaring activity, the formation of hypermassive NS following coalescence of binary neutron stars, etc. 
are capable of exciting quasinormal modes of a pulsar and hence may emit GWs. The timing glitch can be one of 
these sources, which has the potential to excite quasinormal modes in the parent pulsar. 
Although the searches for GWs during August 2006 Vela pulsar (Abadie {\em et al.} 2011) timing 
glitch produced no detectable GWs, with the improving sensitivity of advanced detectors, 
continuous attempts in this direction may produce more conclusive results in near future.

In the literature, there have been a few other theoretical works that advocate 
gravitational wave astronomy in the context of isolated neutron stars. There have been 
discussions on the emission of GWs from deformed neutron stars (Zimmermann \& Szedenits 1979), crustal mountains 
in radio pulsars or in low mass X-ray binaries (Haskell {\em et al.} 2015), continuous emission of GWs from a 
triaxially symmetric rotating neutron star due to permanent ellipticity (Jones 2002), etc. 
Further details on the above-mentioned sources of GWs from isolated pulsars can be found in a detailed 
review by Lasky (2015). In the context of bursts of GWs from an isolated NS, there has also 
been an interesting suggestion by Bagchi {\em et al.} (2015), where authors have proposed 
a unified model for glitches, antiglitches, and generation of GW from isolated pulsars due to phase transitions
inside the core of a pulsar.

In view of these theoretical proposals, we explore here another possible source 
of GWs from isolated pulsars, which may arise due to a very fast-changing elastic deformation 
($\epsilon$) and hence the change in quadrupole moment (Q) of a pulsar as a result of crustquake. 
Nearby Crab pulsar exhibiting small size glitches of order $10^{-8}$ that can be explained successfully 
using the crustquake model will be the most likely candidate to test our proposal. In this context, we 
should mention the work by Keer \& Jones (2015) on crustquake initiated excitation of various oscillation 
modes in a pulsar and hence the emission of possible GWs from such oscillations.
In that work, the authors have assumed that excitations decay after executing several oscillations
and they have made an order of magnitude estimate for the strain amplitude of GWs 
arising due to the excitation of f-mode. According to their picture, the estimated 
strain amplitude depends on various factors such as amplitude of oscillations, 
rotational frequency of the star, etc. in contrast to our picture where $h$ turns out to be 
independent of these quantities.

In this work, we assume that the decay of crustquake initiated excitation behaves like a
critically damped oscillator and estimate the characteristics of GW bursts. 
We will take pulsar of spheroidal (oblate) shape, the ellipticity/oblateness ($\epsilon$) 
of which can be defined through, $I_{zz} \neq I_{xx} = I_{yy}$ and $\epsilon = \frac{I_{zz} - I_{xx}}{I_{0}}$. 
Here, $I_{zz} = \frac{2}{5} M c^2, I_{xx} = \frac{1}{5} M (a^2 + c^2)$ are the MI of the star about (symmetric) z-axis 
and x-axis, respectively ($ c < a$).  $I_0$ is the MI of the spherical star. The quadrupole moment of the spheroidal
star can be taken as, $Q_{zz} = \frac{2}{5} (c^2 - a^2) \equiv Q $ which is related to the ellipticity through, 
$Q = - 2~ I_0 \epsilon$. The negative value of $Q$ arises solely due to oblate shape of the star. Obviously, 
a spheroidal star does not radiate continuous gravitational waves due to its spherical symmetry. However, 
the crustquake can initiate excitations to the star and the star de-excites itself to achieve 
a new equilibrium (more oblate to less oblate) through several oscillations (like an underdamped oscillator) as discussed by Keer \& Jones (2015) 
or it can decay like a critical/overcritical damped oscillator (as we propose in this paper). 
This causes a change in the pulsar's oblateness and hence it should emit gravitational radiation. 
Note that the detailed mechanism of excitation or de-excitations are still unknown, neither the consequences of 
these are established experimentally. In this work, we assume the evolution of ellipticity behaves like, 
$\epsilon(t) \propto exp (-\frac{t}{\tau})$. Where, $\tau$ characterizes the relaxation time scale during 
which major changes of $\Delta \epsilon$ or equivalently,  $\Delta Q$ of QM occurs following crustquake. 
Such an assumption can be justified from the fact that the oscillations caused by excitation is not expected 
to have a well defined single (angular) frequency ($\omega$). Rather, it's more natural to expect that it 
consists of an infinitely large number of normal mode frequencies. The study of de-excitation can then be 
achieved through Fourier analysis by modeling the evolution of ellipticity like a critical/overcritical 
damped oscillators. Also, as the strain amplitude $h(t)$ depends on $\ddot Q^2(t)$,
generation of GWs with a significant strain amplitude is expected only if $Q$ changes in a very short 
duration (i.e., smaller $\tau$) which can be achieved only if the star behaves like a critical/overcritical
damped oscillators. It'll be interesting to study the properties  of matter in details to check if at all a star 
behaves like such damped oscillator.

The motivation for theoretical studies on possible emission of GWs from isolated pulsars can be manifold. 
Firstly, the event rate of (number of events which can produce significant GWs in a galaxy per year) GWs 
emission from pulsars is expected to be quite high in comparison to the event 
rate from various other sources of gravitational radiation, namely mergers of neutron stars (NS) and/or 
black holes (BH). For example, it has been mentioned in Ref. Riles (2013) that event rates lie 
in the  range $2 \times 10^{-4}$ - $ 0.2$ per year for initial LIGO detection of a NS-NS coalescence, and 
$7 \times 10^{-5}$ to 0.1 per year for a NS-BH coalescence, etc. In contrast, there is a huge catalogue of pulsars 
and more than 450 glitch events (crustquake can account for small size $10 ^{-8} - 10^{-12}$ glitches) have been 
recorded \footnote{http://www.jb.man.ac.uk/pulsar/glitches/gTable.html} and analyzed (Ezpinoza {\em et al.} 2011) 
quite extensively. Thus crustquake can be quite frequent phenomenon when one considers a very large number of pulsars 
in our galaxy. The other purpose of the GWs study from pulsar might be to understand a few features of the 
pulsar itself. For example, in this proposal, the strain amplitude of GWs crucially depends on the relaxation time scale ($\tau$) towards the new equilibrium value of oblateness (hence, QM). Thus gravitational wave studies from pulsars can shed light on crustquake, possible mechanism of strain relaxation, etc. To the best of our knowledge, so far no serious  attempt has been made to determine strain relaxation time scale. This is because of the fact that in  the context of crustquake model for glitches, $\tau$ is not relevant in determining glitch size 
and other features of glitches. However, as we show below, $\tau$ is crucial to have a significant value of $h$ 
in our model. Similarly, if a few small size glitches are caused by crustquake, then the glitches will be 
followed by little bursts of GWs can be a falsifiable prediction of a crustquake.

The paper is organized in the following manner. In section 2, we briefly review the basic relevant features of crustquake model as pictured by Baym \& Pines (1971). We shall also discuss here the maximum possible values of ellipticity parameter ($\epsilon$) of a neutron star crust which can sustain the crustal stress. We'll present the methodology in section 3. Here, we shall provide the expressions of GW strain amplitude $h(t)$, characteristic time scale $\tau$, characteristic signal amplitude $h_c(f)$ and the signal to noise ratio (SNR). The results will be presented in section 4, followed by our concluding remarks in section 5.

\section{Crustquake : The Basic Relevant Features}
Crustquake was originally proposed by Ruderman (1968) as a possible explanation for the sudden spin-up
(glitches) of otherwise highly periodic pulsar PSR 0833-45 (Vela). Currently, the superfluid vortex model 
(Anderson \& Itoh 1975) is the leading model to explain glitches.  
However, in the context of glitches or otherwise, crustquake model has been revisited repeatedly. There have been 
discussions in the literature suggesting the involvement of crustquake in NS physics, such as an explanation for the 
giant magnetic flare activities observed in magnetars (Thompson \& Duncan 1995; Lander {\em et al.} 2015), a trigger mechanism for 
angular momentum transfer by vortices (Eichler \& Shaisultanov 2010; Melatos {\em et al.} 2008; Warszawski \& Melatos 2008) or to explain 
the change in spin down rate that continues to exist after a glitch event (Alpar {\em et al.} 1994). In view of these 
discussions, crustquake seems to be an inevitable event for a large number of NS. This work will assume the 
occurrence of crustquake in a NS and study one of it's many possible consequences, namely, the generation of 
short duration ({\it bursts}) GWs due to crustquake.

The crustquake in a neutron star is caused due to the existence of a solid elastic 
(Ruderman 1969; Smoluchowski \& Welch 1970)
deformed crust of thickness about $10^5$ cm. The deformation parameter of the crust can be characterized by its 
ellipticity, $\epsilon = \frac{I_{zz} - I_{xx}}{I_{0}}$ as defined earlier in Section I. Here, we briefly discuss 
the basic features of crustquake model (Baym \& Pines 1971). At an early stage of formation, the crust solidified with 
initial oblateness $\epsilon_o$ (unstrained value or reference value) at a much higher rotational frequency. 
As the star slows down, the ellipticity $\epsilon (t)$ decreases, leading to the development of strain in the 
crust due to the inherent crustal rigidity.  Finally, once the breaking stress is reached, the crust cracks and 
releases its (partial) energy. 
The total energy of the pulsar can be written as (Baym \& Pines 1971), 
\begin{equation} \label{eq:energy}
E = E_0 + \frac{L^2}{2I} + A\epsilon^2 + B(\epsilon- \epsilon_0)^2 .
\end{equation}
Where A and B are two coefficients, the values of which are available (Baym \& Pines 1971). The equilibrium value
of $\epsilon$ can be obtained by minimizing $E$ while keeping angular momentum ($L=I\Omega$) fixed. The value thus obtained 
is given by,
\begin{equation} \label{eq:epf}
\epsilon = \frac{\Omega^2}{4(A+B)}\frac{\delta I}{\delta\epsilon} + \frac{B}{A+B}\epsilon_0 \simeq \frac{B}{A+B}\epsilon_0 .
\end{equation}
The crustquake causes the shift in the reference value of $\epsilon_0$ by $\Delta \epsilon_0$, which in turns changes the
equilibrium value $\epsilon$ by $\Delta \epsilon$, where
\begin{equation} \label{eq:eq-ref}
\Delta\epsilon = \frac{B}{A+B} \Delta\epsilon_0 .
\end{equation}
For normal neutron star of 10 km radius and 1 km crust thickness, we have $A \simeq 6 \times 10^{52}
~ {\rm erg}$ and $B \simeq 6 \times 10^{47} ~{\rm erg}$ (Baym \& Pines 1971) which in turn provides the value 
of $\epsilon$ as, 
\begin{equation} \label{eq:ee0}
\epsilon = 10^{-5} \epsilon_0 .
\end{equation}
The upper limit of ellipticity can be constrained by noting that the crust has a critical strain, say, $\Delta_{cr}$ such
that, 
\begin{equation} \label{eq:strain}
|\epsilon - \epsilon_0| = \frac {A}{B} \epsilon < \Delta_{cr}.
\end{equation}
The theoretical value of $\Delta_{cr}$ ($10^{-2} - 10^{-4}$) had been estimated earlier by Jones (2002). 
However, in a recent work (Horowitz \& Kadau 2009) on crustal breaking strain 
of NS, the authors have done detailed molecular dynamics simulations by modeling the crust by single pure crystal 
and obtained the value, $\Delta_{cr} = 0.1$. 
As per their work, the value of $\Delta_{cr}$ remains around 0.1 even in the presence of impurities, defects, etc. 
Note, this value is at least an order of magnitude higher than the maximum value quoted in Ref. Jones (2002). 
Here for the estimate of maximum GWs strain amplitude, we take $\Delta_{cr} = 0.1$ and we obtain 
the upper limit of $\epsilon$ as, 
\begin{equation} \label{eq:epsilon}
\epsilon < \frac {B}{A} \Delta_{cr} = 10^{-6}.
\end{equation}
At the onset of crustquake, the above value will be taken as an initial oblateness, $\epsilon_i =10^{-6}$ for the 
calculation of GW strain amplitudes. We will assume the change of ellipticity,  $\Delta \epsilon = \eta \epsilon$ 
(where, $0 < \eta < 1$ is a numerical factor characterizing the fraction of strain released due to crustquake.) to 
estimate strain relaxation time $\tau$. The above estimate has been provided for normal NS. It has been proposed that 
more exotic quark star (see the Ref. Keer \& Jones (2015) and the references therein) may exist having a very large 
solid core which can sustain a large ellipticity. For such a star, $A \simeq 8 \times 10^{52}~{\rm erg}$ \& $B 
\simeq 8\times 10^{50}~{\rm erg}$ and hence $\epsilon_i$ can be as large as $10^{-3}$ (for the same value of 
$\Delta_{cr} = 0.1$ as above) as obtained from Eq. \ref{eq:epsilon}. Thus, the quark star can also be a potential 
source of gravitational wave burst as a result of crustquake.

\section {Gravitational Wave Burst Caused by Crustquake : Methodology}
As per the discussion in sec.1, the strain amplitude due to change of 
ellipticity following starquake will be determined by assuming that the star behaves like a 
critically (Now onwards we'll do the analysis for critically damped oscillator. However, the methodology
will be equally applicable for overcritical) damped oscillator. We will take the evolution of ellipticity as, 
\begin{equation}
\epsilon(t) = \epsilon_{i} ~ e^{-t/\tau} = \epsilon_i ~e^{-at} ; (1/\tau \equiv a).
\end{equation}
Here,  $\epsilon_{i}$ is the initial value of ellipticity of the neutron star at the onset
of crustquake. As we are interested to study the evolution of $h(t)$ within characteristic 
time $\tau$, we are neglecting $e^{-at}*t$ term in the above expression. Expanding $\epsilon(t)$ 
in a Fourier series,
\begin{equation}
\epsilon(t) = \frac{1}{\sqrt{\pi}} \int_{0}^{\infty} d\omega~\epsilon(\omega) ~e^{-i\omega t}. 
\end{equation}
Here, $\epsilon(\omega)$ is the corresponding (complex) amplitude in frequency domain and given by, 
\begin{equation}
\epsilon(\omega) = \frac{1}{\sqrt{\pi}} \int_{0}^{\infty} dt~\epsilon(t) ~e^{i\omega t} = 
\frac{\epsilon_i}{\sqrt{\pi}} \Bigg(\frac{a + i\omega}{a^2 + \omega^2}\Bigg).
\end{equation}
Hence, the quadrupole moment can be written as,
\begin{equation} \label{eq:im-QM}
Q(t) = -2 ~I_{0} ~\epsilon(t) = - \frac{2 ~I_0 \epsilon _i}{\pi} \int_{0}^{\infty} 
\frac{a+i\omega}{a^2 + \omega^2} ~ e^{-i\omega t} d\omega .
\end{equation}
Expansion of $\epsilon(t)$ and/or $Q(t)$ in a Fourier series in frequency domain 
allows us to apply linear perturbation theory for the estimate of strain amplitude 
from a source oscillating with infinitely large number of normal modes.
The resultant strain amplitudes $h_{ij}(t)$ can be obtained by summing up the
contribution from each frequency. The strain amplitude (i.e., one of the components say, 
$h_{zz}(t) = h(t)$) in a particular frequency interval can then be written as Riles (2013),
\begin{equation}\label{eq:h01}
h(t) = \frac{2G}{c^4 d} \ddot{Q} = \frac{4GI_0}{\pi c^4 d} \int_{\omega_{min}}^{\omega_{max}}
d\omega ~\epsilon(\omega) ~\omega ^2  e^{-i\omega t} .
\end{equation}
Where, $d$ is the distance of the pulsar from earth and we have taken a frequency interval between 
$\omega_{min}$, $\omega_{max}$ which are the minimum and maximum angular frequencies respectively.
The upper cut-off frequency is a requirement to validate the slow motion approximation. 
In such approximation, the wavelength $\lambda$ of gravitational radiation must be much larger than the 
size of the source. Thus a frequency of about 1 kHz implies the wavelength of radiation as about $\lambda = 300 $ km,  
which is quite large compared to the size of the source ($\sim$ 10 km). We will set 1 kHz frequency as an 
upper limit and slow motion approximation will be assumed upto this frequency. Error arising due to small 
deviation from this approximation for a one kHz frequency will be neglected. There is no such requirement 
on minimum limit of frequency, however, we will estimate $h$ in the frequency range 100 Hz - 1 kHz in 
accordance with the sensitivity of existing interferometers. Thus the strain amplitude [the real part of 
Eq. (\ref{eq:h01})] can be written as,
\begin{equation}
h(t) = \frac{4GI_0\epsilon_i}{\pi c^4 d} \int_{\omega_{min}}^{\omega_{max}}
\Bigg [\frac{a~cos (\omega t) + \omega~sin (\omega t)}{a^2 + \omega^2}\Bigg]~\omega^2~d\omega .
\end{equation}
Expressing in terms of dimensionless quantity $\omega^\prime = \frac {\omega}{a} \equiv \omega \tau$,
the above expression can be re-written as, 
\begin{eqnarray} \label{eq:h02}
h(t) &=& \frac{4GI_0\epsilon_i}{\pi c^4 d~ \tau^2} \int_{\omega^\prime_{min}}^{\omega^\prime_{max}}
\Bigg [\frac{cos (\frac{\omega^\prime t}{\tau}) + \omega^\prime~sin (\frac{\omega^\prime t}{\tau})}
{1 + {\omega^\prime}^2}\Bigg]~{\omega^\prime}^2~d\omega^\prime \nonumber \\
&=& 3.5 \times 10^{-24} \Bigg(\frac{1 {\rm kpc}}{d}\Bigg) \Bigg(\frac{\epsilon_i}{10^{-6}}\Bigg)
\Bigg(\frac{10^{-4} {\rm sec}}{\tau}\Bigg)^2~K(t) .
\end{eqnarray}
The numerical prefactor is calculated by taking MI of the star $I_0$ as $10^{45}~ {\rm gm-cm^2}$. The
function $K(t)$ (for a fixed set of values of $\omega^\prime_{min}$ and $\omega^\prime_{max}$) is given by, 
\begin{equation} \label{eq:ft}
K(t) = \int_{\omega^\prime_{min}}^{\omega^\prime_{max}}
\Bigg [\frac{cos (\frac{\omega^\prime t}{\tau}) + \omega^\prime~sin (\frac{\omega^\prime t}{\tau})}
{1 + {\omega^\prime}^2}\Bigg]~{\omega^\prime}^2~d\omega^\prime .
\end{equation}
The frequency dependent information in the burst can also be inferred from the frequency density 
distribution, $h_{\omega^\prime}(t) = \frac{dh(t)}{d\omega^\prime}$ and is given by 
(using Eq. (\ref{eq:h02})),  
\begin{equation} \label{eq:h-freq}
h_{\omega^\prime}(t) = N \Bigg [\frac{cos (\frac{\omega^\prime t}{\tau}) + 
\omega^\prime~sin (\frac{\omega^\prime t}{\tau})} {1 + {\omega^\prime}^2}\Bigg]~{\omega^\prime}^2 .
\end{equation}
Note, $h_{\omega^\prime}(t)$ is a dimensionless quantity, to be multiplied by $\tau$ to
get the proper frequency distribution $h_\omega(t)$. The prefactor ($N$) in the above equation is of 
order $10^{-24}$ ( to be discussed in the next section). Now, as the strain amplitude depends 
on the relaxation time $\tau$, we will provide a rough estimate for $\tau$ as follows. Here we should 
mention that, so far no so serious attempt has been made to determine the precise value of $\tau$ 
(which is a typical glitch time). It is understood, since in the context of crustquake model for 
glitches, this is not necessary as the size of glitches and other relevant features of glitches are 
not sensitive to the value of $\tau$. If crustquake causes glitches, then uncertainty of such time scale 
may be resolved through pulsar timing by narrowing down to the characteristic time-scale for spin up 
events/glitches. At this stage we may take it as a parameter which can be fixed by observation in future
by studying the impact of $\tau$ on various consequences. However, we'll provide a rough estimate of
$\tau$ by assuming that the strain energy goes into gravitational radiation and the emission is dominated 
within time interval $\tau$. Thus the estimate provided below may set a lower limit on $\tau$. With this 
assumption, we can now write the rate of energy loss due to GW radiation as,
\begin{equation}\label{eq:enr-rate}
\frac{dE}{dt} = -\frac{3G}{10c^5} \dddot{Q}^2 = -\frac{6G}{5c^5} ~I_0^2 \epsilon_i^2 ~a^6 ~e^{-2at}.
\end{equation}
Here we have used, $Q_{zz} = - 2 Q_{xx} = -2 Q_{yy}$ relevant for spheroidal pulsar and 
quadrupole moment tensor $Q_{ij}$ is traceless, i.e., $Q_{zz} + Q_{xx} + Q_{yy} = 0 $.
The net energy loss $\Delta E$ in a typical time $\tau$ is now can be obtained from 
Eq. (\ref{eq:enr-rate}) as,
\begin{equation}\label{eq:tau0}
\Delta E = -\frac{6G}{5c^5} ~I_0^2 \epsilon_i^2 ~a^6 \int_{0}^{1/a} dt ~e^{-2at} \simeq \frac{3G}{5c^5} 
~\frac{I_0^2 \epsilon_i^2}{\tau^5} .
\end{equation}
We can estimate the value of the relaxation time scale $\tau$ by (approximately) equating the above energy 
loss with the strain energy released by the star using Eq.(\ref{eq:energy}), 
\begin{equation}
\Delta E = B ~\Delta \epsilon \simeq  \frac{3G}{5c^5} ~\frac{I_0^2 \epsilon_i^2}{\tau^5} .
\end{equation}
i.e., 
\begin{equation}\label{eq:tau}
\tau = 5 \times 10^{-5} ~ \Bigg (\frac{\epsilon_i}{10^{-6}}\Bigg)^{2/5} 
\Bigg (\frac{10^{39}~{\rm erg}}{B \Delta \epsilon}\Bigg)^{1/5} ~{\rm sec}.
\end{equation}
We will use Eq. (\ref{eq:h02}), (\ref{eq:ft}) and (\ref{eq:tau}) to determine the GW strain 
amplitude $h$ by taking the appropriate values of $\epsilon_i$ and $\Delta \epsilon$.
Here we should mention that while calculating rate of energy loss using Eq. (\ref{eq:enr-rate}), 
we have essentially taken contribution of all frequencies. However, for consistency, one should 
apply Eq. (\ref{eq:im-QM}) and carry out the integration over frequency upto a maximum limit 
(say, 1 kHz) to validate slow motion approximation. However, the rate of energy loss being 
$\frac{1}{\tau^5}$ dependent, the order of magnitude of $\tau$ will not be affected significantly
due to the above factor.

\section{Results}
The strain amplitude in Eq. (\ref{eq:h02}) of gravitational waves in a given frequency
interval in our scenario now depends on $\epsilon_i$ and on the time scale $\tau$ of strain relaxation. 
Following the arguments provided in Sec. 2, for a normal NS the maximum possible value of $\epsilon_i$ 
is in the order of $10^{-6}$ and the value of $\Delta \epsilon$ (required to determine $\tau$) will be 
determined by the prefactor $\eta$ ( $0 < \eta < 1 $). The value of pre-factor $\eta$ and hence, 
$\Delta \epsilon$ can be constrained through the crustquake model for glitches. For small size glitches 
in Crab Pulsar,  $\Delta\epsilon = 10^{-8}$ (Espinoza {\em et al.} 2011), which corresponds to $\eta = 0.01$ 
(i.e., $1\%$ release of strain). For consistency we should mention that for a Crab pulsar, 
the above value of $\Delta \epsilon$ corresponds to only few years waiting time for the next glitch 
to occur, which is much less than the spin-down time scale ($\approx 10^3$ years) of Crab pulsar 
and this fact is fairly consistent with the observations. For a quark star, the value of $\epsilon_i$ 
can be of order $10^{-3}$, an enhancement of order $10^3$ times compared to a normal neutron star.

We will take the value of $\tau$ as obtained from Eq.(\ref{eq:tau}). Assuming $B = 6 \times 10^{47}$ erg, 
$\epsilon_i = 10^{-6}$ and $\Delta\epsilon = 10^{-8}$, we obtain the value of $\tau \simeq 10^{-4}$ sec. 
For exotic quark stars, $\tau$ 
being proportional to $(\frac{\epsilon_i^2}{B \Delta \epsilon})^{1/5}$ is of same order as for a normal NS. 
The strain amplitude $h(\tau)$ for Crab as obtained from Eqs. (\ref{eq:h02}) - (\ref{eq:ft}) is provided 
in Table 1. The time evolution of $h(t)$ is shown 
in Fig (1a). Here, $h(t)$ is the frequency integrated (over the frequency range 100 Hz - 1 kHz) strain 
amplitude. It is obvious that shorter the time scale $\tau $ for spin up events, larger the strain amplitude 
one expects. As we see from the table, the value of strain amplitude for Crab is expected to be of order 
$10^{-25}$. For a same distance quark star $h$ ($ = 10^{-22}$) is enhanced by order $10^3$ compared to Crab. 
The frequency density distribution of strain amplitude $\frac{dh}{d\omega^\prime}/h(\tau)$ (dividing by
$h(\tau)$ is just for convenience) evaluated at $t =\tau$ as obtained from Eq.( \ref{eq:h-freq}) is shown 
in Fig(1b). The plot shows that the proper frequency density $\frac{dh}{d\omega}/h(\tau)$ 
(using $\frac{dh}{d\omega} = \tau ~ \frac{dh}{d\omega^\prime}$) is of order $10^{-3}$. 
Thus, the strain amplitude $h(t,\omega)$ is almost frequency independent in the frequency range 100 Hz - 1 kHz.
\begin{figure*}[h]
\centering
\begin{subfigure}[b]{0.49\textwidth}
\includegraphics[width=3.5in,height=2.7in]{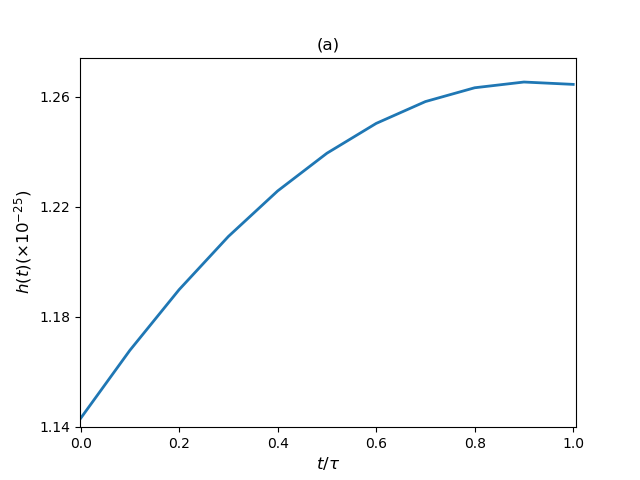}
\centering
\end{subfigure}
\begin{subfigure}[b]{0.49\textwidth}
\includegraphics[width=3.5in,height=2.7in]{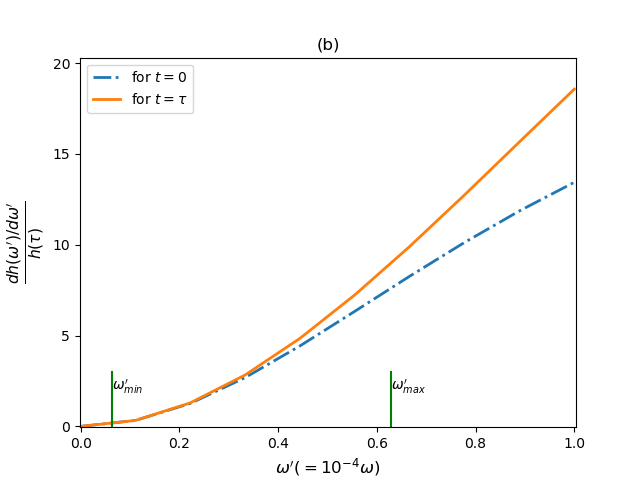}
\centering
\end{subfigure}%
\caption[\footnotesize] {\footnotesize  
(a) Plot shows the time evolution of strain amplitude $h(t)$ within characteristic time, $\tau = 10^{-4}~{\rm sec}$. 
Here, $h(t)$ is the frequency integrated (over the frequency range 100 Hz - 1 kHz) amplitude. 
(b) Plot shows the frequency density distribution of the strain amplitude 
$\frac{dh(\omega^\prime)}{d\omega^\prime}$ as a function of $\omega^\prime$ ($= \omega \tau$). 
$\frac{dh(\omega^\prime)}{d\omega^\prime}$ has been calculated at time $t = \tau$. 
Note, the above distribution is dimensionless and it has to be multiplied by $\tau$ to get the proper
frequency density distribution $\frac{dh(\omega)}{d\omega}$.}
\label{graphs}
\end{figure*}
 
Now whether such bursts of GWs have the potential to be detectable against signal noises depends on various factors 
such as sensitivity of interferometer, proper \lq template' to analyze burst, etc. 
There are several ways to characterize the detectability of signal strength from a source against the 
background noise. One such commonly used method (Sathyaprakash \& Schutz 2009; Moore {\em et al.} 2014) is the comparison of square 
root of power spectral density (PSD) for the source,  $\sqrt {S_h(f)}= \frac {h_c(f)}{\sqrt{f}}$ and for 
the noise, $\sqrt {S_n(f)}= \frac {h_n(f)}{\sqrt{f}}$. Here, $h_c(f)$ is the {\it characteristic signal 
amplitude} in the frequency domain and  $h_n(f)$ is the {\it effective noise} that characterizes the 
sensitivity of a detector. Note, $S_h(f)$ \& $S_n(f)$ both have a dimension of $Hz^{-1/2}$ 
(see Refs. (Sathyaprakash \& Schutz 2009; Moore {\em et al.} 2014) for details), whereas $h_c(f)$ \& $h_n(f)$ are dimensionless. 
The characteristic signal amplitude $h_c(f)$ is defined as, $h_c(f) =  f~|\tilde h_c(f)|$. Where, 
$\tilde h_c(f)$ is the Fourier transform of the signal amplitude $h(t)$. In our case, $\tilde h_c(f)$ 
can be estimated (Sathyaprakash \& Schutz 2009) as, 

\begin{equation}\label{eq:hcf}
\tilde h_c(f) \propto \int_{0}^\infty dt~h(t) e^{i 2\pi f t} \simeq h ~\Bigg(\frac{a+i 2\pi f}{a^2+ 4\pi^2 f^2}\Bigg).
\end{equation}
In the above equation, we have used the fact that the strain amplitude is approximately constant  during the time 
interval $\Delta t = \tau$ ($\frac{\Delta h}{h} \simeq 0.1$ as shown in Fig. (1a)). The characteristic 
signal amplitude is then given by,
\begin{equation}
h_c (f) = |f~ \tilde h_c(f)| = h \frac{f}{\sqrt{a^2 + 4\pi^2f^2}}.
\end{equation}
For $\tau = 10^{-4}$ sec. and for a frequency range 100 Hz - 1 kHz, the characteristic signal strain lies in the range 
$h_c(f) = (0.01 - 0.1)~h$. Hence, the signal to noise ratio, $SNR = \frac{h_c(f)}{h_n(f)}$ can be estimated using the
data provided for the noise amplitude $h_n(f)$. For $h_n(f)$, we utilize the data available in the 
literature (Sathyaprakash \& Schutz 2009; Hild \& Abernathy 2011) for the Einstein Telescope 
\footnote{refer http://www.et-gw.eu/index.php/etsensitivities\#references} and the values lie 
between $10^{-23} - 10^{-22}$ for the frequency range 100 Hz - 1 kHz. For these values of $h_n(f)$, 
SNR (at $f$ = 1 kHz) in our case turns out be of order $10^{-4}$ for Crab and $0.1$ for a same distance quark star.

\begin{table}
\centering
\caption{Typical order of magnitude for strain amplitude, $h$ and characteristic signal amplitude, $h_c(f)$ for 
the bursts. The values of $h$ are provided for Crab and a near-by hypothetical quark star at $t = \tau$ 
for $\tau = 10^{-4}$ sec. The values of $\epsilon_i$ are chosen as per the argument provided in the text.
Characteristic signal amplitude $h_c(f)$ is quoted for a frequency of 1 kHz.}
\label{tab:example_table}
\begin{tabular}{lccc} 
		\hline
		 Star & $\epsilon_i$ &  $h (t=\tau)$ & $h_c(f =  1~{\rm kHz})$  \\		
		\hline\\
		Crab & $10^{-6}$ & $10^{-25}$ & $10^{-26}$ \\
		(d = 2 kpc) & &  &    \\
		\hline 
		Quark star  & $10^{-3}$  &  $10^{-22}$ & $10^{-23}$  \\
		(d = 2 kpc) & &  & \\    
		\hline
\end{tabular}
\end{table}

We conclude this section by comparing strain amplitudes as obtained from isolated pulsars 
(through different mechanism) which were suggested by several authors earlier. For example, 
as we have already pointed out in section 1 that the authors in Ref. Keer \& Jones (2015) have 
estimated the value of strain amplitude resulting from neutron star oscillations initiated 
due to starquake (through somewhat different model). Their suggested values for a kHz frequency 
are larger by an order of magnitude, however, there was an assumption that energy transfer to 
oscillations is instantaneous (i.e., causality is not taken into consideration). 

Similarly, GW emission from a triaxial pulsar has been discussed in the literatures (Jones 2002). 
Although the triaxiality is not a criterion in our scenario, however, even a triaxial star may undergo 
crustquake. In this case, the continuous emission of GWs will be followed by sudden emission of GW burst. 
The strain amplitudes from a triaxial Crab pulsar turns out to be of order 
$(h)_{tr}~\sim 10^{-26}$ for the values of ellipticity, $\epsilon_{tr} = 10 ^{-6}$. The 
above quoted values of $(h)_{tr}$ is provided for Crab (time period T = 33 ms). As $(h)_{tr}$ depends on time 
period of rotation, the magnitude can be even smaller for slowly rotating NS. Our estimate of strain amplitude 
for the burst of GW as a result of crustquake (refer table \ref{tab:example_table}), turns out to be of 
an order of magnitude larger compared to the corresponding value for triaxial stars. We should
emphasize that GW bursts in our case is expected even for a spheroidal star for which continuous emission
will be completely absent. 

Also, there is a proposal by Bagchi {\em et al.} (2015) that several phase transitions inside the core of a pulsar 
can generate quadrupole moment which can result in emission of GW bursts. Our estimated strain 
amplitude turns out to be smaller than the values as obtained in Ref. Bagchi {\em et al.} (2015). This is mainly 
due to smaller value of characteristic time scale $\tau$ (about one microsecond, typical value for a 
QCD phase transition time). We have also presented our estimate of $h$ for a typical quark star (assumed to be 
at a same distance as Crab pulsar). The strain amplitude for such exotic star (about $10^{-22}$) is 
increased by about $10^3$ order of magnitude compared to a normal star. This happens due to high 
ellipticity $\epsilon_i$ at the onset of crustquake. 

\section{Conclusion}
We investigated a possibility of generation of gravitational wave burst caused by crustquake  
due to a very fast change of QM of the star. Modeling the decay of crustquake initiated excitation 
behaves like a critically damped system, we estimated the possible values of strain amplitudes of 
the burst. As we have mentioned, such modeling (i.e., exponential decay) can be justified as 
the excitation expected to have a very large number of frequencies instead of a well defined single 
frequency. With this model, GW strain amplitude depends on ellipticity $\epsilon_i$ at the onset of 
crustquake and relaxation time scale $\tau$. We fixed the value of $\epsilon_i$ ($= 10^{-6}$) as 
suggested in Ref. Horowitz \& Kadau (2009) based on detailed molecular dynamics simulations on NS crust. 
For $\tau$, we have provided an order of magnitude estimate ($\simeq 10^{-4}$) by assuming the
strain energy goes into emission of GWs. For this set of values, $h(\tau)$ (integrated over 
frequency range 100 Hz - 1 kHz) turns out to be of order $10^{-25}$ for Crab and the value is almost 
constant within time $\tau$. We also estimated the characteristic signal amplitude $h_c(f)$ and the 
signal to noise ratio (SNR) for the burst. We noticed that there is a multifold increase in the values 
of these quantities for an exotic quark star. We again point out that the value of $h(t)$
depends on $\tau$, the precise value of which is still unknown. It's therefore important to 
fix this characteristic time scale (which is a typical spin-up time of glitch) through observation 
to reduce the uncertainty in estimating strain amplitude in this model. 
So far, the best resolved time for spin-up has been reported (Dodson {\em et al.} 2002) to be $\approx$ 40 sec. 
for Vela pulsar, which is far away from the requirement. We are hopeful that next generation telescopes 
such as MeerKAT, Giant Magellan, Square Kilometer Array, etc. will enhance the chances of direct glitch 
(sort of {\it as it happens}) detections and constrain the characteristic spin-up time for pulsar glitches.

As far as detectability of gravitational waves is concerned, it is more challenging to detect short duration bursts than continuous GWs.
We hope such challenges will be overcome through proper analysis suitable 
for burst study and eventually by detecting with more sensitive advanced detectors. If so, the study of 
gravitational wave astronomy of pulsars can shed light on possible occurrence of crustquake event itself.
Finally, although we have restricted our study to crustquake (for the purpose of fixing various parameters 
from observation or otherwise), this analysis has the potential to be used for the study of any excitations which 
might be caused by other mechanisms in the context of pulsars. 

\section*{Acknowledgements} 
Pradeepkumar Yadav would like to thank D.I. Jones and Garvin Yim for useful discussion. We also
thank the anonymous reviewers for critical comments and suggestions on our earlier manuscript.

\end{document}